# A Comparison of Partitioning Strategies in AC Optimal Power Flow

Alexander Murray, Michael Kyesswa, Philipp Schmurr, Hüseyin K Çakmak, Veit Hagenmeyer
Institute for Automation and Applied Informatics
Karlsruhe Institute of Technology
Karlsruhe, Germany
{alexander.murray, michael.kyesswa, hueseyin.cakmak, veit.hagenmeyer}@kit.edu

*Abstract*—The highly non-convex AC optimal power flow problem is known to scale very poorly with respect to the number of lines and buses. To achieve improved computational speed and scalability, we apply a distributed optimization algorithm, the so-called Augmented Lagrangian based Alternating Direction Inexact Newton (ALADIN) method. However, the question of how to optimally partition a power grid for use in distributed optimization remains open in the literature. In the present paper, we compare four partitioning strategies using the standard IEEE 9, 14, 30, 39, 57, 118, 300 bus models as the performance benchmark. To this end, we apply the graph partitioners KaFFPa and METIS as well as two spectral clustering methods to the aforementioned bus models. For larger grids, KaFFPa yields the best results on average.

## I. Introduction

Today, human life relies on electricity in almost every facet of life. To ensure reliable supply of power and system stability, Optimal Power Flow (OPF) can be used to derive system settings and operation limits. OPF is typically cast as a numerical optimization problem which seeks to minimize the cost of power generation while ensuring continuous grid operation. A survey of OPF is given in [1] and [2]. Power grids are typically represented as graphs, where the network nodes and branches are the graph vertices and edges, respectively.

In recent studies, parallel and distributed computing approaches are shown to offer great potential for computational speed up and efficiency in grid simulations [3, 4, 5, 6]. The approaches used to parallelize and distribute the optimal power flow problems mainly rely on spatial decomposition as applied in [6, 7]. However, efficient parallel and distributed solutions first require appropriate partitioning of the power grid. Most graph partitioning approaches are hierarchical and split a graph into two partitions per recursion level [8, 9]. Such an approach is called bi-partitioning, and recursive bi-partitioning can create a k-partitioning, but it is not a computationally efficient means of doing so [10]. While not computationally efficient, recursive bi-partitioning has been shown to make more efficient use of memory resources [11]. Some specialized partitioning methods have been developed for use in distributed optimal power flow, such as the spectral method of [9]. Such an approach seeks to create partitions based on information gleaned from the Hessian of the Lagrangrian of the OPF problem, however this requires a priori knowledge of the optimal operating point. Furthermore, both bi-partitioning and spectral clustering are known to scale poorly and other techniques, such as multilevel graph partitioning (MGP), are required for large graphs [12].

Distributed optimization algorithms typically are applied to large-scale problems where runtime, memory, or communication bandwidth is an issue. In the present paper, a variety of partitioning strategies are applied to a variety of IEEE test cases. The well-known Augmented Lagrangian Based Alternating Direction Inexact Newton (ALADIN) algorithm [13] is then used on these partitioned grids to obtain the optimal set-point in a distributed framework and observe the impact of proper spatial partitioning on computational efficiency. The main contribution of the present paper is thus a proof of concept and evaluation of state-of-the-art power grid partitioning strategies.

The rest of this paper is organized as follows: Section II describes the formulation of the OPF problem, and is followed in Section III by a description of the ALADIN distributed optimization algorithm, and the partitioning methods in Section IV. In section V, results from the partitioned cases are described regarding the speedup and efficiency. Section V presents the discussion regarding the performance and proposition of how this performance can be improved in future implementations.

## II. Problem Formulation

The AC optimal power flow problem (AC-OPF) is a constrained non-convex optimization problem that involves some form of Kirchoff's voltage and current laws [1, 2]. While OPF is sometimes used to refer to any problem with this particular structure, herein we will simply consider AC-OPF problems where the objective to be minimized is the generator cost, whose form is given in (1). The box constraints for the decision variables are given in (2), and the AC power flow constraints are shown in (3) and (4).

$$f(x) = \sum_{i \in G} f_i(x) \tag{1}$$

$$\text{with } f_i(x) = a_i \cdot p_i^2 + b_i \cdot p_i + c_i$$

$$\underline{p_i} \leq p_i \leq \overline{p_i}, \quad \underline{q_i} \leq q_i \leq \overline{q_i}, \quad \underline{v_i} \leq v_i \leq \overline{v_i} \tag{2}$$

$$v_i \cdot \sum_{k \in N_i} v_k \cdot (g_{ik} \cdot cos(\theta_{ik}) + b_{ik} \cdot sin(\theta_{ik})) = p_i - p_{d_i} \tag{3}$$

$$v_i \cdot \sum_{k \in N_i} v_k \cdot (g_{ik} \cdot sin(\theta_{ik}) - b_{ik} \cdot cos(\theta_{ik})) = q_i - q_{d_i} \tag{4}$$

Where $x$ is the vector of the physical state variables of the power system. This vector includes the voltage angle $\theta_{ik}$ between nodes $i$ and $k$, the voltage magnitude $v_i$, active injected power $p_i$ and reactive injected power $q_i$ for every bus $i$ within the power system. $p_i$ and $q_i$ are zero for every bus without a generator. For every bus $i$ the constant vectors $p_{d_i}$ and $q_{d_i}$ denote active and reactive power demands, respectively. The constants $g_{ik}$ and $b_{ik}$ correspond to the conductance and susceptance of the branch between bus $i$ and its neighbor $k$. All generator buses are combined in the set $G$ and $f_i(x)$ is the cost function for a single generator bus. It contains three cost parameters $a_i$, $b_i$ and $c_i$ which are defined for every generator. Equations (3) and (4) represent the active and reactive power flow manifold, and can be derived from Kirchoff's laws. Finally, note that as voltage angle and magnitude are relative, one of the buses needs to be defined as the reference bus, or slack bus. The slack bus has the equality constraints $\theta_r = 0$ and $v_r = const$.

## III. OPTIMIZATION METHOD

### A. Distributed Optimization Method

While many algorithms have been developed for non-linear programs (NLPs) in a centralized setting, much less are available which fit a distributed computing context. The ALADIN algorithm is one such method that also guarantees convergence to local optimality for non-convex NLPs [13]. As the AC-power flow manifold constitutes a highly non-convex constraint in the AC-OPF problem, ALADIN is a natural choice of distributed optimization algorithm. Furthermore, it should also be noted that recent work has shown faster convergence for the AC-OPF compared to other distributed optimization methods, like ADMM [14].

The steps of the ALADIN algorithm are outlined in Algorithm 1. The algorithm is initialized with an initial guess of the primal solution $z^0$, an initial vector of the dual variable of the QP $\lambda^0$, the penalty parameters $\rho$ and $\mu$ as well as the weighting matrix $\Sigma_i$. The algorithm works iteratively until the coupling constraint violation and the relative change per iteration is smaller than a given $\epsilon$.

---

**Algorithm 1: ALADIN**

**Input:** Initial guess $x_0 \in X$, $\lambda > 0$, $\rho > 0$, $\mu > 0$, $\Sigma \succeq 0$ and a numerical tolerance $\varepsilon > 0$.

**Initialization:** Set $y = x_0$.

1. Solve for all $i \in \{1, \ldots, N\}$ the decoupled NLPs
$$x_i^* = \min_{x_i} f_i(x_i) + \lambda^T A_i x_i + \frac{\rho}{2} \|x_i - y_i\|_{\Sigma_i}^2$$
$$\text{s.t.} \quad h_i(x_i) \leq 0 \mid \kappa_i$$

2. Compute local gradients, Hessians, and active sets for QP
$$g_i = \nabla_{x_i} f_i(x_i)|_{x_i = x_i^*}$$
$$H_i = \nabla_{x_i}^2 (f_i(x_i) + \kappa_i^T h_i(x_i))|_{x_i = x_i^*}$$
$$C_{i,j}^* = \begin{cases} \nabla_{x_i}(h_i(x_i))_j|_{x_i = x_i^*} & \text{if } (h_i(x_i^*))_j = 0 \\ 0 & \text{otherwise} \end{cases}$$

3. Solve coupling QP
$$\Delta x^* = \operatorname*{argmin}_{\Delta x, s} \frac{1}{2} \Delta x^\top H \Delta x + g^\top \Delta x + \lambda^\top s + \frac{\mu}{2}\|s\|_2^2$$
$$\text{s.t.} \quad A(x + \Delta x) - s = b \mid \lambda_{QP}$$
$$C^* \Delta x = 0$$

4. Line search (optional)
Implement according to Algorithm 3 in [13] to obtain $\alpha_1, \alpha_2, \alpha_3$,
otherwise let $\alpha_1 = \alpha_2 = \alpha_3 = 1$

5. Termination check
If $\|Ax^* - b\|_1 > \delta$ **and** $\sum_{i=1}^{n} \rho \|\Sigma_i(x_i^* - y_i)\|_1 > \delta$, terminate.
Otherwise, update $y \leftarrow y + \alpha_1(x^* - y) + \alpha_2 \Delta x^*$, $\lambda \leftarrow \lambda + \alpha_3(\lambda_{QP} - \lambda)$, and go to Step 1.

ALGORITHM 1: ALADIN ALGORITHM OF [13]

---

In the first step, the uncoupled sub-problems are solved. Note that the local sub-problems do not use their original objective function but rather the so-called augmented Lagrangian; a crucial requirement for ALADIN's convergence properties. Semantically put, the updated objective function allows the algorithm to modify the local sub-problems in a way that it can control the progress towards a feasible solution of the original problem.

After solving the local subproblems in Step 1, a quadratic approximation of the objective function at the current solution, with linearized active sets $C^*$ is constructed in Step 2. The quadratic approximation with coupling constraints is then solved as a centralized QP in Step 3. Note the subproblem solved in Step 3 is an equality constrained quadratic program and thus is solvable, under certain regularity conditions, as a linear system of equations via the linear-independence Karush-Kuhn-Tucker conditions. In Step 4, an optional line search may be performed. Doing so is sometimes helpful for certain problems, but in many cases is unnecessary. Finally, the primal and dual variables are updated according to the solutions of the Steps 1 and 3. The algorithm then proceeds to the next iteration and continues until the termination criteria are met.

### B. Network Partitioning

To apply ALADIN to the optimal power flow problem, it must first be put into a separable, but linearly coupled form. The original problem definition needs to be partitioned into subproblems with appropriate auxiliary variables and linear coupling constraints added at the interfaces between each subproblem. Figure 1 demonstrates this process.

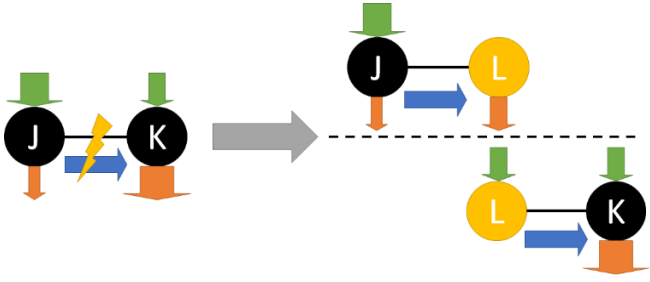

FIGURE 1: CONSTRUCTION OF AN AUXILIARY BUS FOR A CUT-BRANCH.

The cut-branch between buses J and K gets split and an auxiliary bus L is inserted in both partitions (separated by the dashed line). The colored arrows represent power flowing in the given direction and their width represents the amount of power. Loads are colored orange, injection is colored green and power transferred on a branch is colored blue. This results in the following coupling constraints between the introduced auxiliary variables: voltage magnitude and the respective angle need to be identical in both partitions, while the active and reactive power need to be the additive inverse. This can be seen in Figure 1 with the green and orange arrows, which indicate power inflow and outflow, respectively. The introduction of auxiliary buses and equality constraints means that although the new problem is mathematically equivalent to the centralized version of OPF, different algorithmic behavior will emerge. This is due largely to the fact that an initialization of the primal and dual variables associated with the auxiliary buses must be made. This choice will inevitably affect the number of iterations until convergence of the algorithm, and the number of added auxiliary variables will affect the runtime.

## IV. PARTITIONING ALGORITHMS

### A. KAFFPa

The main criterion for speed is the number of branches that connect partitions, where fewer connections typically result in better performance. A secondary criterion is the equality of partition sizes, as the result gathering takes as long as the slowest parallel task. These two criteria perfectly fit the common partitioning problem in graph theory described by Buluç in [12]. In the present paper, the "Karlsruhe Fast Flow Partitioner" (KaFFPa) from the "Karlsruhe High Quality Partitioning" (KaHIP) project is used to generate equally sized partitions that have a minimal number of cut branches. As no a priori knowledge of the optimal operating point is assumed, all power grids given as input shall be considered to have equally weighted branches. Therefore, a minimal total cut results in a minimal number of branches that cross partitions.

KaFFPa consists of three main steps: contraction, initial partitioning, and refinement [15, 16]. In the first step, the algorithm contracts the input graph to create a smaller representation until it is small enough to be partitioned with a global algorithm. To find edges to contract, the algorithm creates a maximum matching using the global paths algorithm (GPA) which was presented in [17]. A matching is a set of edges where no two edges in the set have a common endpoint (vertex). Different weight functions influencing the construction of these matchings are evaluated by Holtgrewe et al. [18]. A matching is then contracted by combining the start and endpoint of every edge in the set. This contraction process quickly decreases the size of the input graph. As soon as the graph is small enough, a global partitioning algorithm is applied. Once finished, the contraction is then undone step by step, applying local refinement strategies. Essentially, this involves checking whether moving some vertices that lie at the partition boundary to the neighboring partition would improve the partition balance or the minimum number of edges cut. A detailed description of the KaFFPa algorithm is given in [15] and [19]. The result is a partitioning with evenly sized partitions and a minimal number of edges that are cut in between the partitions.

### B. Spectral Clustering

The spectral clustering partitioning approach is based on the consideration of the affinity between elements in the system to be partitioned. In the optimal power flow application considered in this paper, this corresponds to the connectedness between the system nodes. The partitioning is therefore performed based on the definition of an affinity matrix and groups the nodes according to their corresponding levels of connectedness.

For purposes of evaluating the performance of the KaFFPa partitioning strategy, two spectral clustering algorithms are applied in the present paper to produce additional partitions. In the rest of the present paper, the approaches are referred to as "*Spectral clustering – Ybus*" and "*Spectral clustering – Hessian*". The difference between the two algorithms is how the affinity matrix is defined. In the *Spectral clustering – Ybus* approach, the affinity matrix is formed using the admittance matrix according to the algorithm described in [8]. The *Spectral clustering – Hessian* approach combines the Hessian matrix of the optimization problem and the admittance matrix to form the affinity matrix, as proposed in [9].

### C. METIS

METIS partitioning is used for partitioning large unstructured graphs. The approach uses undirected graphs as inputs, which are then split into the required number of subsystems [20]. Like KaFFPa, it is a multi-level partitioning method where the main requirement is an equal number of nodes in each subsystem and a minimum number of interconnections between the partitions, however it uses a Kernigan-Lin approach in the uncoarsening phase as opposed to the local search method used by KaFFPa. Overall, the runtime is still comparable to KaFFPa.

## V. NUMERICAL RESULTS

The results in this section are obtained using ALADIN as the distributed optimization algorithm applied to partitionings of the standard IEEE test networks as implemented in MATPOWER [21]. The partitionings are generated by the algorithms described in Section IV. Shown in Figure 2 are the 2-partitions generated by all strategies for the IEEE 14 bus case. These partitions are in line with the intuitive partitioning of these grids.

Shown in Table 1 are the number of iterations required by Algorithm 1 to converge to a locally optimal solution when applied to each of the partitioned OPF problems. Each of the local subproblems are solved with IPOPT [22] and the equality constrained QPs are solved as a linear system via the KKT-conditions. In all cases, the termination threshold is $10^{-3}$. An iteration limit of 500 is set for Algorithm 1, which is reached in the 300 bus case for several partitionings.

| Case - partitions | KAFFPa | Spectral Clustering Hessian | Spectral Clustering Ybus | METIS |
|---|---|---|---|---|
| 9 - 2 | 3 | 3 | 3 | 3 |
| 14 - 2 | 17 | 29 | 29 | 17 |
| 30 - 2 | 20 | 15 | 15 | 23 |
| 39 - 2 | 4 | 8 | 8 | 4 |
| 57 - 2 | 16 | 14 | 20 | 41 |
| 57 - 3 | 22 | 24 | 29 | 71 |
| 118 - 2 | 20 | 23 | 23 | 22 |
| 118 - 3 | 28 | 34 | 40 | 37 |
| 118 - 4 | 28 | 29 | 25 | 22 |
| 118 - 5 | 23 | 42 | 44 | 44 |
| 300 - 3 | 51 | 28 | >500 | 102 |
| 300 - 5 | 97 | >500 | 54 | 103 |

TABLE 1 : ALADIN ITERATIONS FOR PARTITIONED IEEE TEST CASES

| Case - partitions | KAFFPa | Spectral Clustering Hessian | Spectral Clustering Ybus | METIS |
|---|---|---|---|---|
| 9 - 2 | (500, 500, 1.05, 2) | (500, 500, 1.05, 2) | (500, 500, 1.05, 2) | (500, 500, 1.05, 2) |
| 14 - 2 | (500, 500, 1.05, 2) | (500, 500, 1.05, 2) | (500, 500, 1.05, 2) | (500, 500, 1.05, 2) |
| 30 - 2 | (500, 1000, 1.05, 2) | (500, 500, 1.05, 2) | (500, 500, 1.05, 2) | (500, 1000, 1.05, 2) |
| 39 - 2 | (20, 2000, 1.15, 1.15) | (500, 2000, 1.15, 2) | (500, 2000, 1.15, 2) | (500, 2000, 1.15, 2) |
| 57 - 2 | (1000, 2000, 1.05, 2) | (1000, 2000, 1.05, 2) | (1000, 2000, 1.05, 2) | (100, 100, 1.2, 2) |
| 57 - 3 | (100, 2000, 1.2, 2) | (100, 2000, 1.2, 2) | (100, 2000, 1.2, 2) | (500, 2500, 1.05, 1.15) |
| 118 - 2 | (100, 100, 1.2, 2) | (100, 100, 1.2, 2) | (100, 100, 1.2, 2) | (100, 100, 1.2, 2) |
| 118 - 3 | (100, 1000, 1.05, 1.5) | (100, 1000, 1.05, 1.5) | (100, 1000, 1.05, 1.5) | (100, 1000, 1.05, 1.5) |
| 118 - 4 | (500, 1000, 1.05, 1.5) | (500, 1000, 1.2, 1.5) | (500, 1000, 1.1, 1.5) | (500, 1000, 1.1, 2) |
| 118 - 5 | (500, 2000, 1.1, 2) | (500, 2000, 1.1, 1.2) | (500, 2000, 1.1, 1.2) | (500, 2000, 1.1, 1.2) |
| 300 - 3 | (100, 100, 1.1, 2) | (100, 100, 1.1, 2) | N/A | (100,100,1.05,1.1) |
| 300 - 5 | (100, 100, 1.1, 2) | N/A | (100, 100, 1.1, 2) | (500, 500, 1.05, 2) |

TABLE 2 : BEST ALADIN PARAMETERS FOR EACH PARTITIONING

As can be seen in Algorithm 1, ALADIN requires certain parameters to be given, which not only can be difficult to choose a priori, but also have a large effect on the convergence rate. A number of different combinations of parameters are tested for each of the partitionings and only the best results are shown in Table 1. Interestingly, the best parameters seem to vary from partitioning to partitioning. Shown in Table 2 are the parameters used to generate the results of Table 1. All parameters are given in the form ($\rho,\mu,\dot{\rho},\dot{\mu}$) where $\rho$ and $\mu$ are the initial values of the parameters described in Section III, and where $\dot{\rho}$ and $\dot{\mu}$ are the factors by which $\rho$ and $\mu$ are updated at every iteration. It should be noted that this updating method is a heuristic means of replacing the line search step.

While there is some variability, the results shown in Table 1 are relatively good if one considers a worst-case partitioning strategy. For a worst-case partitioning, we use the singleton partitioning, where every bus is in its own partition. While it may seem like there is a lot of potential for parallelization with such an approach, it actually leads to a significantly larger problem due to the introduction of so many auxiliary buses and a very large QP to solve in Step 3 of Algorithm 1. The results of this partitioning are shown in Table 3. For the IEEE 57, 118, and 300 bus cases there were no parameters found which allowed for ALADIN to converge within 500 iterations.

| Case | Iterations | ALADIN parameters |
|---|---|---|
| 9 | 3 | (500, 500, 1.05, 2) |
| 14 | 42 | (500, 500, 1.05, 2) |
| 30 | 162 | (50, 250, 1.05, 1.05) |
| 39 | 4 | (20, 2000, 1.15, 1.15) |
| 57 | >500 | N/A |
| 118 | >500 | N/A |
| 300 | >500 | N/A |

TABLE 3: RESULTS FOR THE SINGLETON PARTITIONING

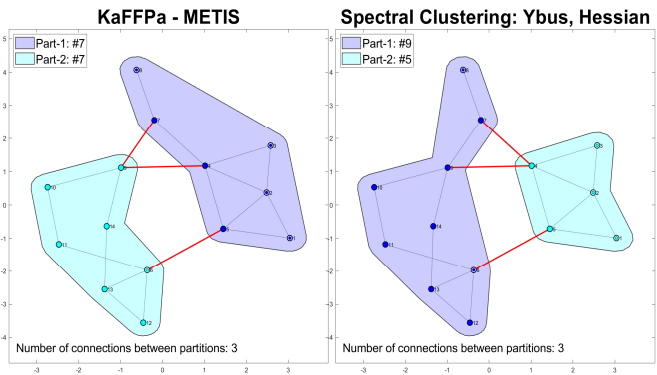

FIGURE 2: PARTITIONS GENERATED FOR THE IEEE 9 AND 14 BUS CASES.

The IEEE 57 bus case is the smallest example examined such that all of the partitioning algorithms provided different partitionings. Shown in Figure 3 are the partitionings of each algorithm. Observe that although the partitions generated by KaFFPa and METIS are quite similar, there is nonetheless a large difference in the convergence rate of ALADIN. As seen in Figure 3, METIS generates many more branches between partitions than KaFFPa, which is likely the reason for its relatively poor performance in the results of Table 1. Also of note is the fact that Spectral Clustering applied to the Y-bus matrix consistently yields almost the exact same partitions as

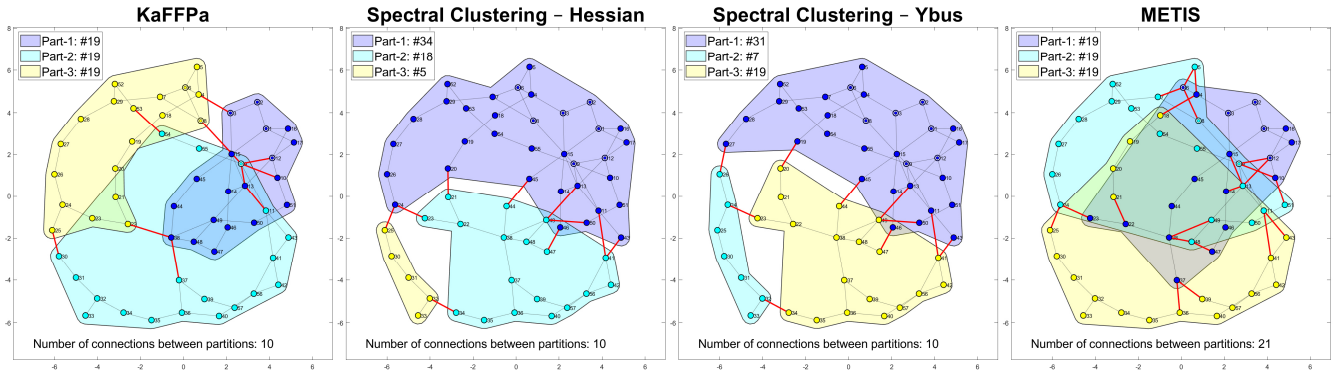

FIGURE 3: PARTITIONS GENERATED FOR THE IEEE 57 BUS CASE. RED LINES INDICATE WHICH BRANCHES CROSS BETWEEN PARTITIONS.

Spectral Clustering of the Hessian of the Lagrangian at the optimal operating point. This seems to imply that the resistances and susceptances of the lines are dominant when considering a partitioning and that very little is gained by the a priori knowledge of the optimal operating point. However, as noted in [9], this may change when line limits are reached.

## VI. CONCLUSIONS

The results presented in this paper have shown that proper problem partitioning can have a large impact on the convergence rate of the applied distributed optimization algorithm. While the specific partitioning algorithm did not seem to have a significant impact on the convergence rate for the smaller cases, there was a difference observed for the larger (118 and 300 bus) cases, wherein KaFFPa generated partitions which led to faster convergence. While these results are highly dependent on the specific initialization strategies and algorithmic parameter choices, this nonetheless indicates that effective problem partitioning can be carried out without first computing an optimal operating point. Furthermore, while the focus of this paper is on the examination of iterations until convergence, KAFFPa and METIS also generate equally-sized partitionings, which affects the time per iteration of ALADIN since the OPF subproblems are equally sized and with few interconnections, and thus a small QP coupling problem. This is an increasingly important consideration as the problem size grows since the disparity in the partition sizes generated by Spectral Clustering can be quite large.

Of note is that some recent research has shown that the coupling QP of ALADIN can be solved efficiently in a decentralized manner [23], however such results are preliminary and their implementation has been left to future work. Other upcoming work will investigate edge weighting methods for the graphs given as input to KaFFPa. Doing so can potentially further improve the partitions generated and lead to more balanced subproblems for the distributed optimization algorithm to solve. Finally, it is worth considering whether optimal power grid partitioning is really required at all. Future work will examine dynamic problem partitioning, wherein partitions are adjusted within each ALADIN iteration. As the spectral clustering method requires almost the exact information that is computed in Step 2 of ALADIN, it may be possible to develop a variant of ALADIN that maintains its convergence and optimality guarantees but has the ability to dynamically adjust a given partitioning in order to improve performance. Such an algorithm would have the advantage of not having to rely on an optimally partitioned system to be given as input.

## VII. REFERENCES

[1] S. Frank, I. Steponavice and S. Rebennack, "Optimal power flow: A bibliographic survey, I: Formulations and deterministic methods," *Energy Syst.*, vol. 3, pp. 221-258, 2012.

[2] S. Frank, I. Steponavice and S. Rebennack, "Optimal power flow: A bibliographic survey, II: Nondeterministic and hybrid methods," *Energy Syst.*, vol. 3, pp. 259-289, 2013.

[3] F. Pellegrini, "Distillating knowledge about SCOTCH," in *Combinatorial Scientific Computing*, 2009.

[4] M.A.Tomim, J.R.Martí and L.Wang, "Parallel solution of large power system networks using the Multi-Area Thévenin Equivalents (MATE) algorithm," *International Journal of Electrical Power & Energy Systems,* vol. 31, no. 9, pp. 497-503, Oct 2009.

[5] J. Liu, M. Benosman and A. U. Raghunathan, "Consensus-based distributed optimal power flow algorithm," in *2015 IEEE Power Energy Society Innovative Smart Grid Technologies Conference (ISGT)*, Washington, DC, USA, Feb 2015.

[6] A. Murray, A. Engelmann and V. Hagenmeyer, "Hierarchical Distributed Mixed-Integer Optimization for Reactive Power Dispatch," in *10th Symposium on Control of Power and Energy Systems*, 2018.

[7] S. H. Low, "Convex relaxation of optimal power flow—Part II: Exactness," *IEEE Transactions on Control of Network Systems,* vol. 1, no. 2, pp. 177 - 189, May 2014.

[8] A. Y. Ng, M. I. Jordan and Y. Weiss, "On spectral clustering: analysis and an algorithm," in *14th International Conference on Neural Information Processing Systems: Natural and Synthetic*, 2001.

[9] J. Guo, G. Hug and O. K. Tonguz, "Intelligent Partitioning in Distributed Optimization of Electric Power Systems," *IEEE Trans. Smart Grid,* vol. 7, no. 3, pp. 1249-1258, May 2016.

[10] S. Schlag, V. Henne, T. Heuer, H. Meyerhenke, P. Sanders and C. Schulz, "k-way Hypergraph Partitioning via n-Level Recursive Bisection," *2016 Proceedings of the Eighteenth Workshop on Algorithm Engineering and Experiments (ALENEX),* pp. 53--67, 2016.

[11] R. Drechsler, W. Gunther, T. Eschbach, L. Linhard and G. Angst, "Recursive bi-partitioning of netlists for large number of partitions," *Euromicro Symposium on Digital System Design, 2002. Proceedings.,* pp. 38--44, 2002.

[12] A. Buluç, H. Meyerhenke, I. Safro, P. Sanders and C. Schulz, "Recent Advances in Graph Partitioning.," in *Algorithm Engineering*, Cham, Springer, 2016, pp. 117-158.

[13] B. Houska, J. Frasch and M. Diehl, "An Augmented Lagrangian Based Algorithm for Distributed NonConvex Optimization," *SIAM J. Optim,* vol. 26, no. 2, p. 1101–1127, Jan 2016.

[14] A. Engelmann, T. Mühlpfordt, Y. Jiang, B. Houska and T. Faulwasser, "Distributed AC Optimal Power Flow using ALADIN," *IFAC-PapersOnLine,* vol. 50, no. 1, pp. 5536 - 5541, 2017.

[15] P. Sanders and C. Schulz, "High quality graph partitioning," in *10th DIMACS implementation challenge workshop: Graph Partitioning and Graph Clustering*, 2013.

[16] P. Sanders and C. Schulz, "Think Locally, Act Globally: Highly Balanced Graph Partitioning," in *Experimental Algorithms*, Springer Berlin Heidelberg, 2013, pp. 164 - 175.

[17] J. Maue and P. Sanders, "Engineering Algorithms for Approximate Weighted Matching," in *International Workshop on Experimental and Efficient Algorithms*, 2007.

[18] M. Holtgrewe, P. Sanders and C. Schulz, "Engineering a scalable high quality graph partitioner," in *2010 IEEE International Symposium on Parallel & Distributed Processing (IPDPS)*, Atlanta, GA, USA, April 2010.

[19] P. Sanders and C. Schulz, "Engineering Multilevel Graph Partitioning Algorithms," in *19th European Symposium on Algorithms (ESA'11)*, 2011.

[20] G. Karypis, "METIS: A software package for partitioning unstructured graphs, partitioning meshes, and computing fill-reducing orderings of sparse matrices Version 5.1.0," University of Minnesota, Department of Computer Science & Engineering, Minneapolis,, March 2013.

[21] R. D. Zimmerman, C. E. Murillo-Sanchez and R. J. Thomas, "MATPOWER: Steady-State Operations, Planning, and Analysis Tools for Power Systems Research and Education," *IEEE Transactions on Power Systems,* vol. 26, no. 1, pp. 12--19, 2011.

[22] A. B. L. Wächter, "On the implementation of an interior-point filter line-search algorithm for large-scale nonlinear programming," *Mathematical Programming,* vol. 106, no. 1, pp. 25-57, 2005.

[23] A. Engelmann, Y. Jiang, B. Houska and T. Faulwasser, "Decomposition of non-convex optimization via bi-level distributed ALADIN," *Transactions on Control of Network Systems ,* vol. (Submitted. Preprint available at arxiv.org/abs/1903.11280 ), 2019.

[24] M. Hill and M. Marty, "Amdahl's law in the multicore era," *Computer,* vol. 41, no. 7, pp. 33--38, 2008.